# Millimeter Wave Path Loss for Diverse Antenna Patterns in Outdoor Environment

Jarosław Wojtuń*†, Cezary Ziółkowski*, Jan M. Kelner*, Paweł Skokowski*, Niraj Narayan※, Rajeev Shukla※, Aniruddha Chandra※, Radek Závorka‡, Tomáš Mikulášek‡, Jiří Blumenstein‡, Ondřej Zelený‡, and Aleš Prokeš‡

*Institute of Communications Systems, Faculty of Electronics, Military University of Technology, 00-908 Warsaw, Poland
※Department of Electronics and Communication Engineering, National Institute of Technology Durgapur, 713209 Durgapur, India
‡Department of Radio Electronics, Brno University of Technology, 61600 Brno, Czech Republic
†jaroslaw.wojtun@wat.edu.pl

*Abstract*—Empirical path loss models are defined for a specific antenna system used during measurements and characterized by a particular radiation pattern and main lobe beam width. In this paper, we propose a novel approach to modifying such a model to estimate path loss for antenna systems with different radiation patterns and beam widths. This method is based on a multi-elliptical propagation model, enabling a more flexible adaptation of the path loss model. The paper presents the general concept of the proposed method and numerical study results demonstrating the influence of the antenna pattern shape and its beam width on path loss estimation.

## I. Introduction

The development of communication technologies and the growing demand for high-quality broadband services have led to the intensive utilization of the millimeter-wave (mmWave) spectrum in fifth-generation (5G) systems. This spectrum, known as frequency range 2 (FR2), i.e., 24.25–71 GHz, has enabled the allocation of significantly wider channels than traditional sub-6 GHz bands, i.e., frequency range 1 (FR1) 410 MHz–7.125 GHz, which in turn allows for higher transmission speeds and better support for advanced applications requiring low latency. As a result, mmWaves play a crucial role in the implementation of high-capacity networks and the simultaneous servicing of a large number of devices [1], [2].

However, mmWave propagation encounters significant challenges related to high signal attenuation and susceptibility to interference and environmental obstacles. Compared to FR1 bands, mmWaves have a shorter range and experience stronger propagation losses. To mitigate these negative effects, mmWave systems employ advanced antenna arrays with narrow beams and high directional gains, e.g., massive multiple-input-multiple-output (MIMO) [3]. This enables efficient transmission over relatively short distances, making mmWaves particularly well-suited for densely deployed pico- and femtocells, forming ultra-dense networks (UDN) [4], [5].

To effectively leverage the capabilities of antenna systems in the FR2 range, precise radio wave propagation modeling is crucial. Traditional models, such as those presented in the 3rd Generation Partnership Project (3GPP) Technical Report (TR) no. 38.901 [6], are initially designed for omnidirectional antennas, necessitating additional algorithms to account for directional antennas. However, implementing these algorithms involves high computational costs. An alternative approach is to use empirical propagation models derived from measurements, which can better reflect real-world propagation conditions. However, these models are typically developed for specific antenna systems [7], [8].

In this paper, we propose a novel method for determining a directional path loss (PL) model for any beam width and shape of an antenna pattern based on an omnidirectional PL model. This innovative approach to modeling propagation losses for mmWave bands is based on a numerical calculation and characterizes by low computational requirements in relation to the simulation solutions proposed by 3GPP TR 38.901 [6].

The input factor for PL calculations is the omnidirectional PL model. It may be obtained based on measurements in a real environment using a narrow-beam directional antenna and then synthesized [8], [9]. Another approach may be to use omnidirectional PL models proposed by the 3GPP standard [6]. In the paper, we adapt the second one.

A key aspect of this modification involves transformations based on a multi-elliptical propagation model (MPM) [10], [11], [12]. We shortly describe how MPM-based numerical calculations determine the modified PL. Then, we show the adaptation of the proposed solution to calculate the directional PL for diverse antenna pattern models and varied half-power beam widths (HPBWs). These analyses are presented for two mmWaves bands, i.e., 28 and 39 GHz, and outdoor urban environment, considering both line-of-sight (LOS) and non-LOS (NLOS) conditions.

The rest of the paper is organized as follows. Section II shows different approach to modeling radiation power pattern of antenna systems. Methodology of determination directional PL model based on omnidirectional one and the MPM is described in Section III. Exemplary results and their analysis, we show in Section IV. Finally, Section V contains the paper summary.

## II. Antenna Pattern Models

A power radiation pattern is one of the antenna system's basic characteristics. The pattern is a three-dimensional (3D) characteristic defined in the elevation and azimuth planes. Our research uses a simplified, two-dimensional (2D) MPM. Therefore, in the rest of the paper, we analyze the pattern only in the azimuth plane (i.e., for an elevation of 90°).

The actual antenna pattern is measured in an anechoic chamber. Such a pattern is usually not symmetrical, which results from the imperfections of the antenna system elements. In practice, we can indicate symmetries in the pattern apart from some small deviations. Therefore, simplified pattern models are usually used instead of real patterns for calculations.

This research was funded in part by the National Science Center (NCN), Poland, grant no. 2021/43/I/ST7/03294 (MubaMilWave). For this purpose of Open Access, the author has applied a CC-BY public copyright license to any Author Accepted Manuscript (AAM) version arising from this submission.

The simplest models, such as Gaussian, consider only main lobes of antenna patterns. While, more complex ones, such as Sinc, also allow for considering the side lobes of antenna patterns [13]. For more complex antenna systems, the use of the mentioned models may sometimes be insufficient. In such cases, more complex patterns should be used.

In our research, we additionally use a realistic pattern model of the massive MIMO system, i.e., a vertical patch as an antenna array of $12 \times 8$ elements. The main lobe HPBWs of the BS antenna beam are 12.6° and 6° for the azimuth (i.e., horizontal) and elevation (i.e., vertical) planes, respectively. The UE beam with HPBWs equal to 90° and 65° for the azimuth and elevation planes, respectively, is generated by a single antenna element. A realistic pattern for such a 5G base station (gNodeB) antenna system is generated based on the methodology recommended by the 3GPP [14] and International Telecommunication Union (ITU) [15]. The user equipment (UE) beam with HPBWs equal to 90° and 65° for the azimuth and elevation planes, respectively, is generated by a single antenna element. The gNodeB, Gaussian, or Sinc antenna pattern models are used as the transmitting antenna. While UE pattern is applied as the receiving antenna. These antenna patterns in the azimuth plane are illustrated in Fig. 1 [13].

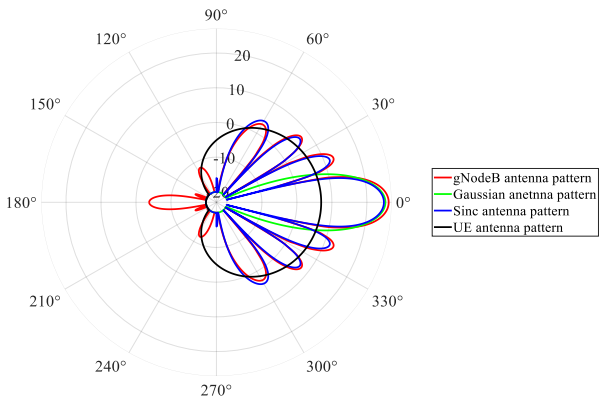

Fig. 1. Power beam antenna pattern models in azimuth plane considered in our studies.

## III. METHODOLOGY OF PATH LOSS CALCULATION

### A. Multi-elliptical Propagation Model

Geometry-based propagation models enable radio signal scattering areas to be modeled using appropriate geometric structures. In practice, various probability distributions are additionally used to differentiate scattering both on a plane and in space, enhancing the realism of this approach. Geometry-based stochastic models (GBSMs) consider both the environment's geometry (e.g., buildings, obstacles) and the statistical properties of radio wave propagation resulting from physical phenomena such as scattering, reflection, and diffraction. This allows for obtaining realistic results with significantly lower computational complexity than deterministic models, such as ray tracing [16].

The ability to adjust the shape and size of geometric structures and the parameters of statistical distributions allows for the adaptation of GBSMs to various environments and propagation conditions. Research findings indicate that these models effectively capture the variability of the radio environment, eliminating the need for precise geometric representation of each object – an essential requirement in deterministic methods. In summary, GBSMs serve as a compromise between simple empirical models and complex deterministic propagation models. Due to their flexibility and computational efficiency, they are widely used in the analysis of radio wave propagation, particularly in modern wireless communication systems [16].

In the modification method, we propose to apply the MPM. In our opinion, the MPM is one of the few GBSMs in which the geometric structure can be directly related to the radio channel transmission characteristics. In this case, a power delay profile (PDP), $P(\tau)$, is used to generate a coaxial multi-elliptical structure, in the foci of which the transmitter (Tx) and receiver (Rx) are located. The size of the ellipse (its major $a_i$ and minor $b_i$ axes, $i = 1,2,...,N$) is defined by the characteristic PDP delays that describe the time-clusters in a radio channel. On the other hand, the powers of these clusters are used for defining the amplitudes of individual paths. In this paper, we use 2D MPM [12]. Generally, you can also consider the elevation plane and then in 3D, we use multi-semi-ellipsoidal geometry structure [10], [11], [12]. Fig. 2 depicts the geometry structure of scattering areas in 2D MPM [12].

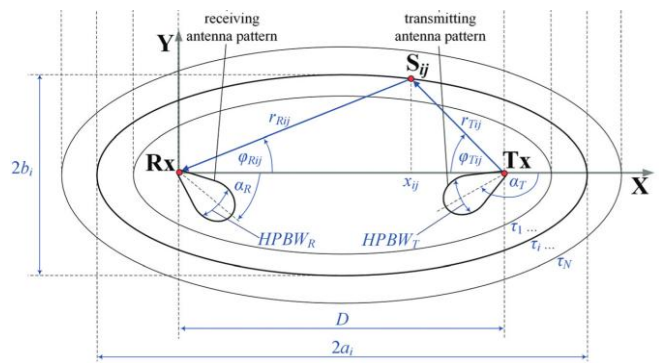

Fig. 2. Geometry structure of scattering areas in MPM.

The MPM is used, i.a., to determine a power azimuth spectrum (PAS), $P(\varphi, D)$, for a specific distance, $D$. In this procedure, we assume that the individual propagation paths are grouped into time-clusters resulting from the PDP. For paths with $\tau > 0$, we have so-called delayed scattering components. For NLOS conditions, paths with $\tau = 0$ form the so-called local scattering components. They are modeled by the von Mises distribution, the shape of which is defined by the local scattering intensity coefficient. Under LOS conditions, for $\tau = 0$, there is an additional direct path component, which is defined by the Rician K-factor [10], [11], [12].

### B. Modification of Path Loss Model

Simplistically, the developed PL modification method is based on the following equation in a logarithm scale:

$$PL_{out}(D) = PL_{in}(D) + PL_{corr}(D), \quad (1)$$

where $PL_{in}$ and $PL_{out}$ are so-called the input and output PLs for omnidirectional and directional antenna patterns, respectively, and $PL_{corr}$ means the PL modification coefficient.

To determine the coefficient, we calculate the PASs for the omnidirectional antenna, $P_{in}(\varphi, D)$, and the analyzed directional antenna, $P_{out}(\varphi, D)$, using the MPM. Therefore,

$$PL_{corr}(D) = 10\log_{10}\left(\frac{G_{in}^{Tx}G_{in}^{Rx}}{G_{out}^{Tx}G_{out}^{Rx}} \cdot \frac{\int_{-180°}^{180°} P_{out}(\varphi,D)d\varphi}{\int_{-180°}^{180°} P_{in}(\varphi,D)d\varphi}\right), \quad (2)$$

where $G_{in}^{Tx}$ and $G_{in}^{Rx}$ are the gains of the omnidirectional Tx and Rx antennas, respectively, $G_{out}^{Tx}$ and $G_{out}^{Rx}$ are the gains of the omnidirectional Tx and Rx antennas, respectively.

As we mentioned, to calculate the input PLs for a specific distance range, we use the 3GPP standard propagation model [6]. Next, the PL modification coefficient is determined based on the MPM. We can consider antenna pattern models with any shapes and different HPBWs in the MPM calculation.

A detailed and analytical description of the proposed modification method is presented in [17]. This work also shows the method verification based on empirical measurements for mmWave.

## IV. RESULTS AND ANALYSIS

### A. Assumptions for Simulation Studies

The input PLs for an omnidirectional antenna pattern are determined based on the 3GPP TR 38.901 [6]. We perform calculations for the distance $D$ between the Tx and Rx varies from 60 to 180 m. We additionally consider the intensity coefficient of the local scattering components equal to $\gamma = 60$. The transmitting and receiving antennas are looking at each other, i.e., full beam alignment – $\alpha_T = 180°$ and $\alpha_R = 0°$ (see Fig. 2).

The proposed approach requires the use of PDP to determine the geometric structure of the MPM. For this purpose, we use the tapped-delay line (TDL) models defined in the 3GPP 38.901 [6], i.e., TDL-C and TDL-E for NLOS and LOS conditions, respectively. These normalized TDLs are adapted to selected propagation environment and frequency range by multiplying them by the appropriate delay spread, $\sigma_\tau$. In our study, we choose the normal-delay profile, urban macro (UMa) scenario, and $\sigma_\tau = 266$ ns or $\sigma_\tau = 249$ ns for the carrier frequency, $f_c$, equal to 28 GHz or 39 GHz, respectively. In TDL-E, the Rician K-factor is equal to $\kappa = 22$ dB.

### B. Impact of Antenna Pattern Model

In the first study, we analyzed three different antenna pattern models. The HPBW of the realistic gNodeB pattern model results from the number of antenna array elements (see Section II). Therefore, we considered the same HPBW in the Gaussian and Sinc models. The obtained results for two frequencies, 28 and 39 GHz, are illustrated in Figs. 3 and 4 for LOS and NLOS conditions, respectively.

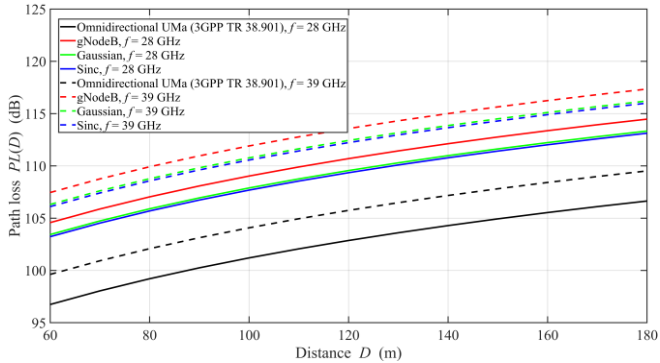

Fig. 3. PL versus distance for diverse antenna patterns under LOS conditions.

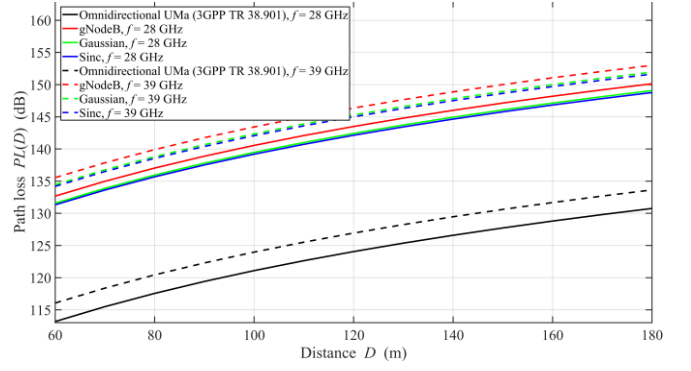

Fig. 4. PL versus distance for diverse antenna patterns under NLOS conditions.

The analysis of the results shows that the PL increases with increasing frequency. The results obtained using the Gaussian and Sinc models are similar, although the first one only considers the main lobe. The PL for gNodeB is slightly larger than for the other two models. Moreover, using directional antennas introduces larger PL in relation to the input omnidirectional antennas. However, it should be remembered that the higher gains of the directional antennas influence on the improvement of the wireless link power budget. On the other hand, introducing an angle-selective beam helps reduce interference and phase differences in the propagation paths reaching the Rx.

### C. Influence of Antenna Beam Width

To determine the impact of the antenna HPBW on PL, we employed our method with the UE and Sinc antenna patterns for the Rx and Tx, respectively. We selected three different HPBWs for the Sinc antenna, namely 8°, 16°, and 24°. In our study, the HPBWs on the azimuth and elevation planes are assumed to be identical. Our values are consistent with real-world antenna parameters. Table I presents typical values for the Ka-band (26–40 GHz) directional horn antenna parameters from Eravant [18]. The results are shown in Figs. 5 and 6 for LOS and NLOS conditions, respectively.

TABLE I. Ka – BAND HORN ANTENNAS PARAMETERS

| Parameters | Antenna [18] | | | | |
| --- | --- | --- | --- | --- | --- |
| | SAR-1532-28-S2 | SAR-1725-28-S2 | SAR-2013-28-S2 | SAR-2309-28-S2 | SAR-2507-28-S2 |
| 3 dB HPBW in elevation plane (°) | 33 | 23 | 14 | 10 | 7 |
| 3 dB HPBW in azimuth plane (°) | 33 | 24 | 16 | 11 | 9 |
| Gain (dBi) | 15 | 17 | 20 | 23 | 25 |

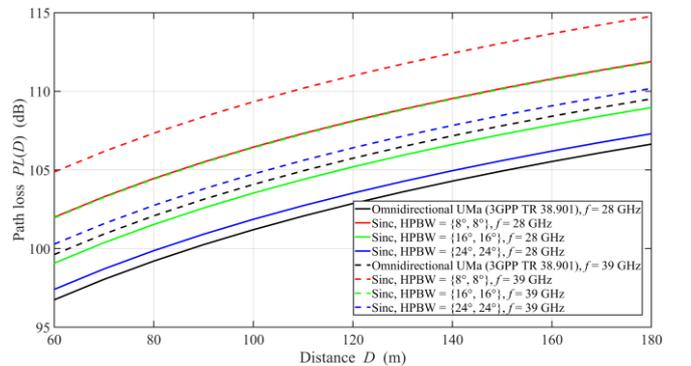

Fig. 5. PL versus distance for different HPBWs under LOS conditions.

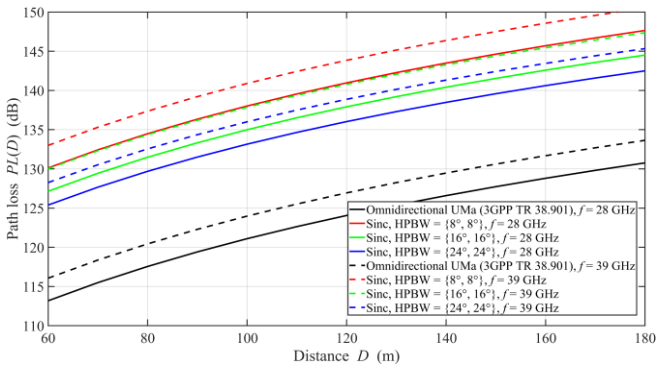

Fig. 6. PL versus distance for different HPBWs under NLOS conditions.

The increase in HPBW causes a decrease in PL. Under LOS conditions, the graphs for HPBW equal to 24° approach those corresponding to directional antennas (see Fig. 5 – black and blue lines). This shows that the PL for antennas with HPBW larger than the analyzed 24° can only slightly increase compared to omnidirectional antennas. This situation results from the fact that the dominant direct path is most important. The scattering delayed components only slightly affect the PL result. For NLOS conditions, this situation looks different. The lack of a direct path causes delayed components to play a crucial role in the PL. In this case, the narrow-beam antenna significantly filters spatially the propagation paths arriving from directions other than the one defined by the main beam lobe.

## V. Summary

This paper presents a novel method for modifying the omnidirectional PL model using MPM-based numerical calculations. Directional PL models defined for different antenna patterns and HPBWs result from the applied modification. The proposed approach is characterized by low complexity and computational time consumption. Moreover, it allows to easily adapt different shapes of antenna patterns. The paper presents exemplary study results for different pattern models and various HPBWs. Moreover, the research considers two mmWave frequencies, LOS and NLOS conditions.

We plan to present a detailed analytical description of the method with the elevation plane (3D) in [17]. In this case, the method will be based on the 3D multi-ellipsoidal propagation model [10], [11], [12]. An empirical verification of the PL modification method and potential areas of its application will also be included in [17].

## Acknowledgment

This work was partly supported by the National Science Centre, Poland, under the OPUS call in the Weave program under research project no. 2021/43/I/ST7/03294 acronym 'MubaMilWave', partly by the Czech Science Foundation under project no. 23-04304L; in part by the Military University of Technology under project no. UGB/22-059/2025/WAT; and partly by the MeitY, Government of India, through the Chips-to-Startup program under project no. EE-9/2/2021-R&D-E.

## References

[1] A. E. C. Redondi, C. Innamorati, S. Gallucci, S. Fiocchi, and F. Matera, "A survey on future millimeter-wave communication applications," *IEEE Access*, vol. 12, pp. 133165–133182, 2024, doi: 10.1109/ACCESS.2024.3438625.

[2] A. N. Uwaechia and N. M. Mahyuddin, "A comprehensive survey on millimeter wave communications for fifth-generation wireless networks: Feasibility and challenges," *IEEE Access*, vol. 8, pp. 62367–62414, 2020, doi: 10.1109/ACCESS.2020.2984204.

[3] S. A. Busari, K. M. S. Huq, S. Mumtaz, L. Dai, and J. Rodriguez, "Millimeter-wave massive MIMO communication for future wireless systems: A survey," *IEEE Commun. Surv. Tutor.*, vol. 20, no. 2, pp. 836–869, 2018, doi: 10.1109/COMST.2017.2787460.

[4] Y. Teng, M. Liu, F. R. Yu, V. C. M. Leung, M. Song, and Y. Zhang, "Resource allocation for ultra-dense networks: A survey, some research issues and challenges," *IEEE Commun. Surv. Tutor.*, vol. 21, no. 3, pp. 2134–2168, 2019, doi: 10.1109/COMST.2018.2867268.

[5] X. Wang *et al.*, "Millimeter wave communication: A comprehensive survey," *IEEE Commun. Surv. Tutor.*, vol. 20, no. 3, pp. 1616–1653, 2018, doi: 10.1109/COMST.2018.2844322.

[6] "5G; Study on channel model for frequencies from 0.5 to 100 GHz (3GPP TR 38.901 version 18.0.0 Release 18)," ETSI/3GPP, ETSI TR 138.901 V18.0.0 (2024-05), May 2024. [Online]. Available: https://www.etsi.org/deliver/etsi_tr/138900_138999/138901/17.01.00_60/tr_138901v170100p.pdf

[7] Y. Xing and T. S. Rappaport, "Millimeter wave and terahertz urban microcell propagation measurements and models," *IEEE Commun. Lett.*, vol. 25, no. 12, pp. 3755–3759, Dec. 2021, doi: 10.1109/LCOMM.2021.3117900.

[8] T. S. Rappaport, G. R. MacCartney, M. K. Samimi, and S. Sun, "Wideband millimeter-wave propagation measurements and channel models for future wireless communication system design," *IEEE Trans. Commun.*, vol. 63, no. 9, pp. 3029–3056, Sep. 2015, doi: 10.1109/TCOMM.2015.2434384.

[9] S. Sun, G. R. MacCartney, M. K. Samimi, and T. S. Rappaport, "Synthesizing omnidirectional antenna patterns, received power and path loss from directional antennas for 5G millimeter-wave communications," in *2015 IEEE Global Communications Conference (GLOBECOM)*, San Diego, CA, USA, Dec. 2015, pp. 1–7. doi: 10.1109/GLOCOM.2015.7417335.

[10] C. Ziółkowski and J. M. Kelner, "Statistical evaluation of the azimuth and elevation angles seen at the output of the receiving antenna," *IEEE Trans. Antennas Propag.*, vol. 66, no. 4, pp. 2165–2169, Apr. 2018, doi: 10.1109/TAP.2018.2796719.

[11] C. Ziółkowski and J. M. Kelner, "Antenna pattern in three-dimensional modelling of the arrival angle in simulation studies of wireless channels," *IET Microw. Antennas Propag.*, vol. 11, no. 6, pp. 898–906, May 2017, doi: 10.1049/iet-map.2016.0591.

[12] J. M. Kelner and C. Ziółkowski, "Multi-elliptical geometry of scatterers in modeling propagation effect at receiver," in *Antennas and wave propagation*, P. Pinho, Ed., London, UK: IntechOpen, 2018, pp. 115–141. doi: 10.5772/intechopen.75142.

[13] J. Wojtuń, C. Ziółkowski, and J. M. Kelner, "Modification of simple antenna pattern models for inter-beam interference assessment in massive multiple-input–multiple-output systems," *Sensors*, vol. 23, no. 22, Art. no. 22, Jan. 2023, doi: 10.3390/s23229022.

[14] "Evolved Universal Terrestrial Radio Access (E-UTRA) and Universal Terrestrial Radio Access (UTRA; Radio Frequency (RF) requirement background for Active Antenna System (AAS) Base Station (BS) (3GPP TR 37.842 version 13.3.0 Release 13)," 3rd Generation Partnership Project (3GPP), Valbonne, France, Tech. Rep. 3GPP TR 37.842 V13.3.0 (2019-12), Release 13, Jan. 2020. Accessed: Feb. 09, 2020. [Online]. Available: https://portal.3gpp.org/desktopmodules/Specifications/SpecificationDetails.aspx?specificationId=2625

[15] "SWG Sharing Studies. Working document on characteristics of terrestrial component of IMT for sharing and compatibility studies in preparation for WRC-23," International Telecommunication Union (ITU), Radiocommunication Study Groups, Geneva, Switzerland, Document 5D/TEMP/228-E, Oct. 2020.

[16] J. Chebil, H. Zormati, and J. B. Taher, "Geometry-based channel modelling for vehicle-to-vehicle communication: A review," *Int. J. Antennas Propag.*, vol. 2021, no. 1, p. 4293266, 2021, doi: 10.1155/2021/4293266.

[17] C. Ziółkowski *et al.*, "Directional path loss prediction based on omnidirectional model for multipath environment," *IEEE Trans. Antennas Propag.*, (under review).

[18] "Eravant," Eravant. Accessed: Mar. 05, 2025. [Online]. Available: https://www.eravant.com/